\documentclass[%
 reprint,
superscriptaddress,
showpacs,preprintnumbers,
 amsmath,amssymb,
 aps,
]{revtex4-1}

\usepackage{graphicx}
\usepackage{dcolumn}
\usepackage{bm}

\usepackage{color}

\begin{document}


\title{Transformational acoustic metamaterials based on pressure gradients}

\author{C. Garc\'{i}a-Meca}
\email{Corresponding author: cargarm2@ntc.upv.es}
\affiliation{Nanophotonics Technology Center, Universitat Polit\`{e}cnica de Val\`{e}ncia, 46022 Valencia, Spain.}

\author{S. Carloni}%
\affiliation{Institute of Theoretical Physics, MFF, Charles University,  V. Hole\v{s}ovi\v{c}k\'{a}ch 2 180 00 Praha 8,  Czech Republic }%

\author{C. Barcel\'{o}}
\affiliation{Instituto de Astrof\'{i}sica de Andaluc\'{i}a (CSIC), Glorieta de la Astronom\'{i}a, 18008 Granada, Spain.}

\author{G. Jannes}
\affiliation{Modelling \& Numerical Simulation Group, Universidad Carlos III de Madrid, Avda. de la Universidad, 30, 28911 Legan\'{e}s (Madrid), Spain}%

\author{J. S\'{a}nchez-Dehesa}
\affiliation{Wave Phenomena Group, Universitat Polit\`{e}cnica de Val\`{e}ncia, 46022 Valencia, Spain.}%

\author{A. Mart\'{i}nez}
\affiliation{Nanophotonics Technology Center, Universitat Polit\`{e}cnica de Val\`{e}ncia, 46022 Valencia, Spain.}%

\date{\today}

\begin{abstract}
We apply a homogenization process to the acoustic velocity potential wave equation. The study of various examples shows that the resulting effective properties are different from those of the homogenized pressure wave equation for the same underlying acoustic parameters. A careful analysis reveals that a given set of inhomogeneous parameters represents an entirely different physical system depending on the considered equation. Our findings unveil a different way of tailoring acoustic properties through gradients of the static pressure. In contrast to standard metafluids based on isobaric composites, this alternative kind of metafluids is suitable for the implementation of transformational devices designed via the velocity potential equation. This includes acoustic systems in a moving background or arising from general space-time transformations. As an example, we design a device able to cloak the acoustic velocity potential.
\end{abstract}

\pacs{43.20.+g}
\maketitle


\section{	Introduction}\label{sec:Intro}
In the last years, transformation acoustics has emerged as an interesting tool for designing devices able to manage sound propagation in novel ways~\cite{CUM07-NJP,CAI07-NJP,CHE07-APL,CUM08-PRL,TOR07-NJP,TOR08-NJP,NOR08-JASA,CHE10-JPD,URZ10-NJP,POP11-PRL,GOK12-JASA,Bergamin}. Recently, the authors of this contribution have proposed a new transformation approach to the design of acoustic metamaterials: Analogue Transformation Acoustics (ATA)~\cite{GAR13-SR,GAR14-WM,GAR14-PNFA}.
ATA overcomes some problems that arise when applying transformational techniques to acoustics by using the transformation properties of an analogue model \cite{BAR11-LRR}, instead of the 4D diffeomorphism invariance employed in transformation optics~\cite{POST} (and not valid for the acoustic equation).  More specifically, starting from the well-known~\cite{UNR81-PRL,BAR11-LRR} formulation of the acoustic equation in terms of the perturbation of the velocity potential $\phi_1$ (we use the notation $\partial_t=\partial/\partial t$ and $\partial_i=\partial/\partial x^i$, with $x^1=x$, $x^2=y$ and $x^3=z$)
\begin{eqnarray}
\label{wave_moving}
&-&\partial_t\left[\rho_0{c_0}^{-2}\left(\partial_t\phi_1 +\mathbf{v}_0\cdot\nabla\phi_1\right)\right] \\ \nonumber
&+&\nabla\cdot\left[\rho_0\nabla\phi_1-\rho_0 c_0^{-2}\left(\partial_t\phi_1 + \mathbf{v}_0\cdot\nabla\phi_1 \right)\mathbf{v}_0\right]=0,
\end{eqnarray}
where $\rho_0$, $c_0$ and $\mathbf{v}_0$ are respectively the background or static mass density, speed of sound and velocity, one notes that this equation can be recast as 
\begin{equation} \label{dAlembertian}
{1\over\sqrt{-g}} \partial_\mu \left( \sqrt{-g} \; g^{\mu\nu} \; \partial_\nu \phi_1 \right)=0,
\end{equation}
which represents the motion of a scalar field in a space-time whose metric is defined as 
\begin{eqnarray}
g^{\mu\nu} = \frac{1}{\rho c}
\left(
\begin{array}{ccc}
-1 & \vdots & -{v}_0^i \\
... & . & ................ \\
-{v}_0^i & \vdots & c_{\rm V}^2\tilde{\gamma}^{ij}-{v}_0^i{v}_0^j
\end{array}
\right).~
\end{eqnarray}
This possibility, which lays at the core of the so-called ``analogue gravity" paradigm, is also the cornerstone of ATA. In particular, one applies the classic transformation technique developed for electromagnetism on the analogue model rather than on the corresponding initial acoustic equation (containing the parameters of the virtual medium), and successively maps back the transformed analogue equation to the acoustic one in order to obtain the new (meta-)medium that implements the desired transformation~\cite{GAR13-SR,GAR14-WM,GAR14-PNFA}. The relativistic analogue model (\ref{dAlembertian}) is constructed in such a way to encode the physical properties of the acoustic systems, but at the same time it is characterized  in terms of tensorial equations in an abstract space-time. Therefore, one can consider any space-time coordinate transformation compatible with the acoustic equation (in the sense of being re-interpretable as a medium). For example, since Eq.\eqref{wave_moving} requires an isotropic medium, all the transformations we can consider are the ones that preserve isotropy. We will address this limitation of ATA (at least partially) in the present work. Let us remark that we use here the word analogue model in the opposite direction than usual in analogue gravity. The relativistic abstract system is here the analogue of the real physical system in the laboratory. 

In Refs.~\onlinecite{GAR13-SR,GAR14-WM,GAR14-PNFA}, ATA has been applied with success to the design of a series of devices (e.g., time cloak, frequency shifter) which are typical of electromagnetic systems and that were not workable in acoustics. In those works, the advantages of ATA over Standard Transformation Acoustics (STA) become evident. For example, it was found that introducing a moving background was required to implement transformations mixing space and time. This feature cannot be considered using simply the pressure wave equation commonly used in STA, because the introduction of a background velocity breaks the form-invariance of the pressure wave equation (even under Galilean coordinate transformations). Therefore velocity potential transformations provide, via the ATA technique, a more general way of building transformational devices. In addition, although purely spatial transformations can be worked with both equations, the prescriptions for the acoustic parameters returned by each equation are different, and the ones given by ATA can be more suitable for the construction of real-world devices.

Like in any other transformation approach, also in the case of ATA the prescribed acoustic parameters are smooth functions of the coordinates and show an anisotropic character. However, the actual construction process of acoustic devices relies on the use of natural materials which only provide a discrete set of isotropic acoustic properties. The problem then arises to connect the theoretical results of ATA and the technological realization of the required acoustic media. To solve this problem we can resort to acoustic metamaterials or metafluids, i.e., fluids made up of various materials with certain acoustic parameters (usually homogeneous and isotropic), which under some given conditions display different effective parameters that depend on those of the constitutive materials and on their shapes~\cite{TOR07-NJP,NOR08-JASA}. One way to obtain these effective parameters is to use homogenization techniques~\cite{Bensoussan,HAS98-CS}. In the case of periodic systems whose physics can be described by differential equations with oscillating coefficients, homogenization allows us to approximate the full equations with equivalent ones containing homogeneous coefficients. As far as acoustic metamaterials are concerned, these techniques have been widely used to homogenize the pressure wave equation~\cite{TOR06-PRL,TOR06-PRB,FAR08-NJP,TOR09-PRB}. It is then reasonable to expect that the same approach can be used to achieve a better link between the ATA prescriptions and real systems. 

In this contribution, we address specifically such connection. In section~\ref{sec:Homogenization}, we describe the application of a two-scale homogenization technique to the velocity potential wave equation and present several examples. Surprisingly, we find that the resulting effective parameters are different from those arising from the homogenized pressure equation. In section~\ref{sec:Analysis} we analyze the origin of this behavior. Our analysis reveals new interesting insights on the way in which the velocity potential equation (and thus ATA) works, as well as an alternative way to construct metafluids based on gradients of the static pressure. Finally, in section~\ref{sec:Cloaking} we use the results of section~\ref{sec:Homogenization} to design a device able to cloak the acoustic velocity potential. Some conclusions are drawn in section~\ref{sec:Conclusions}.

\section{Homogenization of the acoustic equations}\label{sec:Homogenization}
Our initial goal is to homogenize the velocity potential equation. For simplicity, we will start by considering a non-moving background ($\mathbf{v}_0=0$), for which Eq.~\eqref{wave_moving} reads~\cite{BLO46-JASA,UNR81-PRL,PIE90-JASA,BAR11-LRR}
\begin{equation}
-C\partial^2_t\phi_1 + \partial_i\left(a^{ij}\partial_j\phi_1\right) = 0,
\label{eq1}
\end{equation}
with $C=\rho_0c_0^{-2}$ and $a^{ij}=\rho_0\delta^{ij}$. In addition, we would like to compare the resulting effective parameters with the ones arising from the homogenization of the pressure wave equation, which are usually considered in the construction of acoustic metamaterials. Eq.~(\ref{eq1}) also represents the pressure wave equation if we take $C=\rho_0^{-1}c_0^{-2}$ and $a^{ij}=\rho_0^{-1}\delta^{ij}$, and replace $\phi_1$ by the acoustic pressure $p_1$~\cite{LANDAU,BER46-JASA,PIE90-JASA}. We will assume that the composite to be homogenized is periodic in such a way that $\rho_0(\mathbf{x})=\rho_0(\mathbf{x}+\mathbf{NY})$ and $c_0(\mathbf{x})=c_0(\mathbf{x}+\mathbf{NY})$, where $\mathbf{N}=diag(n_1,n_2,n_3)$ is a diagonal matrix with $n_i$ an integer number and $\mathbf{Y}=(Y_1,Y_2,Y_3)^T$ is a constant vector that determines the periodicity in each Cartesian direction. Moreover, we will assume that the coefficients $a^{ij}$ satisfy the ellipticity condition $a^{ij}v_iv_j\ge\alpha|\mathbf{v}|^2$ ($\alpha>0$) for all $\mathbf{v} \in \mathbb{R}^3$. Under such conditions, using a two-scale homogenization approach it can be proven that, in the long wavelength limit, Eq.~(\ref{eq1}) is equivalent to the following one (see appendix~\ref{appendix_1})
\begin{equation}
-\tilde{C} \partial^2_t\phi_1^0 + \tilde{a}^{ij}\frac{\partial^2\phi_1(\mathbf{x})}{\partial x^i\partial x^j}=0,
\label{eq2}
\end{equation}
with constant coefficients given by
\begin{equation}
\tilde{C} = \langle C(\mathbf{x}) \rangle    \quad   , \quad   \tilde{a}^{ij}=\left\langle a^{ij}+a^{ik}\frac{\partial \chi^j(\mathbf{x})}{\partial x^k} \right\rangle, 
\label{eq3}
\end{equation}
where $ \langle \cdot \rangle$ represents spatial averaging over the unit cell and where the functions $\chi^k(\mathbf{x})$ are also $\mathbf{Y}$-periodic and satisfy the so-called cell problem
\begin{equation}
\frac{\partial}{\partial x^i}\left(a^{ij}\frac{\partial \chi^k(\mathbf{x})}{\partial x^j}\right)=-\frac{\partial a^{ik}}{\partial x^i}
\label{eq4}
\end{equation}
As a consequence, we can define the following effective anisotropic densities for the homogenized velocity potential and pressure equations,
\begin{eqnarray}
\label{eq5} \tilde{\rho}^{ij}_{\phi}&=&\left\langle \rho_0\delta^{ij}+\rho_0\delta^{ik}\frac{\partial \chi^j(\mathbf{x})}{\partial x^k} \right\rangle \\ 
\label{eq6} (\tilde{\rho}_{p}^{ij})^{-1}&=&\left\langle \rho^{-1}_0\delta^{ij}+\rho^{-1}_0\delta^{ik}\frac{\partial \chi^j(\mathbf{x})}{\partial x^k} \right\rangle 
\end{eqnarray}
Eq.~\eqref{eq6} is the usual definition employed in the literature. Note that, although we have considered a simplified problem with a non-moving background, the appearance of an effective anisotropic density in the homogenized $\phi$-equation already enables us to extend ATA to anisotropic spatial transformations of the velocity potential (which are physically different from transformations of the pressure~\cite{GAR13-SR,GAR14-WM}). However, there is an issue that deserves further attention; the effective densities arising from the homogenization of the $\phi$-equation and the $p$-equation might be different, as deduced from Eqs.~\eqref{eq5} and~\eqref{eq6}. A similar conclusion can be drawn about the effective speeds of sound.

\begin{figure}
\begin{center}
  \includegraphics[width=8cm]{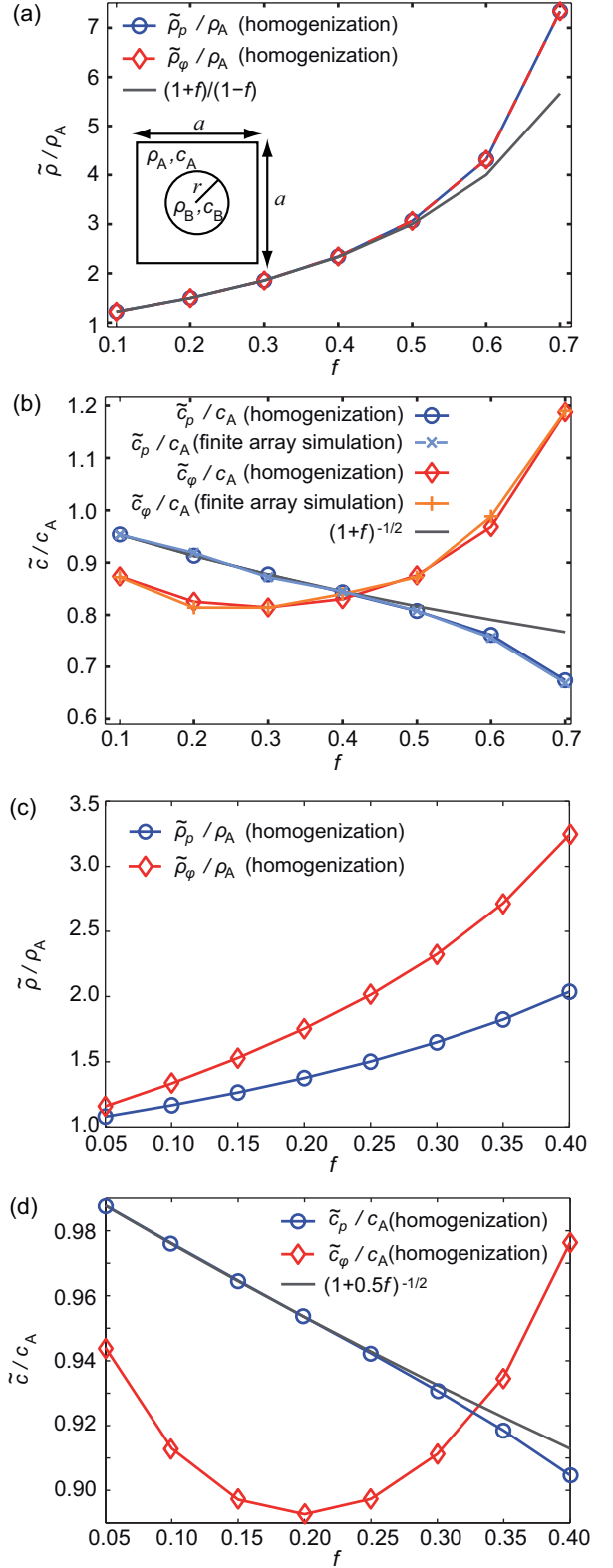}
\end{center}
  \caption{(Color online) Effective parameters obtained for periodic arrays (square lattice with periodicity $l$) of wood inclusions embedded in air as a function of the wood filling fraction $f$ ($f=\pi r^2/l^2$ for infinite cylinders and $f=4\pi r^3/(3l^3)$ for spheres, $r$ being the radius). (a) Effective density and (b) speed of sound of a 2D array of cylinders. The speed of sound is calculated using two different methods (homogenization and simulation of a finite array). (c) Effective density and (d) speed of sound of a 3D array of spheres.}
  \label{Fig1}
\end{figure}

As an example, let us consider the homogenization of a two-dimensional (three-dimensional) periodic array of cylinders (spheres) of radius $r$, with constant background parameters $\rho_B$ and $c_B$ embedded in a fluid with parameters $\rho_A$ and $c_A$. Since the resulting effective media will be isotropic, we can also define the following effective speeds of sound
\begin{eqnarray}
\label{eq7} \tilde{c}_{\phi}^2 &=&\tilde{\rho}_{\phi} \langle \rho_0c_0^{-2} \rangle^{-1}, \\ 
\label{eq8} \tilde{c}_{p}^2 &=& \tilde{\rho}_{p}^{-1} \langle \rho_0^{-1}c_0^{-2} \rangle^{-1},
\end{eqnarray}
where $\tilde{\rho}^{ij}_{\phi}=\tilde{\rho}_{\phi}\delta^{ij}$ and $\tilde{\rho}^{ij}_p=\tilde{\rho}_p\delta^{ij}$. Specifically, let us study a common configuration in which the cylinders (spheres) are made of wood and the surrounding medium is air~\cite{TOR06-PRL}. Thus, we can take $\rho_B\approx 700\rho_A$ and $c_B\approx 10 c_A$. We calculated the corresponding effective parameters by solving numerically Eqs.~\eqref{eq4}-\eqref{eq8} in COMSOL Multiphysics. Note that the derivative on the right-hand side of Eq.~\eqref{eq4} is not defined at the interface between both media. To deal with this situation we solve the corresponding equation at each uniform domain and apply the proper boundary condition at the discontinuity~\cite{FAR08-NJP}. To derive the natural boundary condition associated with Eq.~\eqref{eq4}, particularized for the case $a^{ij} = a_0 \delta^{ij}$, it is wise to express this equation as
\begin{align}
\nabla \cdot \left( a_0 \nabla \chi^i \right) = \nabla \cdot \left(a_0 \mathbf{e}_i \right)
\end{align}
Then, the use of the divergence theorem readily leads us to the following boundary condition
\begin{align}
a_2 \frac{\partial \chi^i}{\partial n} \bigg|_2 - a_1 \frac{\partial \chi^i}{\partial n}\bigg|_1 = -(a_2-a_1)\mathbf{e}_i \cdot \mathbf{n},
\end{align}
where the subscripts 1 and 2 refer to each of the two media($a_1$ and $a_2$ are the values of $a_0$ at medium 1 and 2, respectively), $\mathbf{e}_i$ are the basis vectors in Cartesian coordinates, and $\mathbf{n}$ denotes the unit normal to the boundary between both media, pointing from medium 1 to medium 2. The results for the previous example are shown in Fig.~\ref{Fig1}. 
\par
\begin{figure}
\begin{center}
  \includegraphics{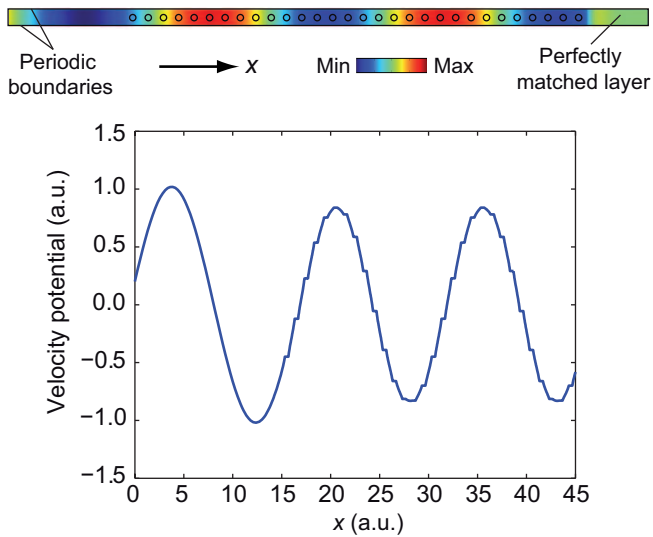}
\end{center}
  \caption{(Color online) Simulation of the acoustic velocity potential in a finite array of wood cylinders embedded in air. A plane wave impinges onto the array from the left. A perfectly matched layer (PML) absorbs the output wave on the right. Periodic conditions are applied at the top and bottom simulation boundaries. Note that a stair-like behavior appears at the cylinder-background interfaces due to the discontinuity in the acoustic parameters.}
  \label{Fig2}
\end{figure}

In the case of the cylinders, the effective density associated with the homogenized $\phi$ and $p$ equations is the same. This should be the expected behavior since, in principle, both quantities represent the same physical system (a system described by the same underlying density and sound speed). As a double-check, we also plot the function $(1-f)/(1+f)$ which is known to provide a very good approximation for the effective density in this kind of system for low filling fractions~\cite{TOR06-PRL}. However, the effective speeds of sound found after the homogenization of each equation are different  [see Fig.~\ref{Fig1}(b)]. We further verified these results by simulating a finite array of wood cylinders in COMSOL for each considered 2D configuration. From the simulation we can retrieve the effective speed of sound inside the finite array, which is in excellent agreement with that predicted by the homogenization process. The velocity potential for one of the simulated finite-array configurations is depicted in Fig.~\ref{Fig2}. In addition, we plot the function $(1-f)^{-1/2}$ in Fig.~\ref{Fig1}(b), a good approximation for the effective speed of sound at low filling fractions. Remarkably, the values of $\tilde{c}$ arising from this approximation only coincide with those associated with the homogenization of the pressure wave equation. 
\par
In the 3D case (wood spheres in air), neither the homogenized densities nor the sound speeds coincide [see Fig.~\ref{Fig1}(c)-(d)]. 
Again, only $\tilde{c}_p$ agrees with the prediction for low filling fractions, given in this case by $(1+0.5f)^{-1/2}$. Clearly, these results must imply that a certain distribution of acoustic parameters $\rho_0$ and $c_0$ represents a different physical system depending on whether it is used in the $\phi$-equation or in the $p$-equation. In section~\ref{sec:Analysis} we look for an explanation to this behavior.

To conclude this section, we briefly analyze a simple yet useful example that we will employ in the construction of a velocity potential cloak. It is that in which the medium only varies in one direction, i.e., $a^{ij}=a_0(x)\delta^{ij}$ ($a_0$ being $Y_1$-periodic). In this case, it can be shown that (see appendix)
\begin{equation}\label{eqa}
\tilde{a}^{ij}=\left(\begin{array}{ccc}
\langle a_0^{-1}(x)\rangle^{-1} & 0 & 0 \\
0 & \langle a_0(x)\rangle & 0 \\
0 & 0 & \langle a_0(x)\rangle
\end{array}\right)\\
\end{equation}
For instance, a periodic multilayer structure made up of two different homogeneous materials falls within the class of systems described by Eq.~\eqref{eqa}. Specifically, if the two materials are characterized by parameters $\rho_{\rm A}$, $c_{\rm A}$ and $\rho_{\rm B}$, $c_{\rm B}$, and the thickness of the layers of each material is $d_{\rm A}$ and $d_{\rm B}$, then
\begin{align} \label{rho_x}
\tilde{\rho}_{\phi}^x&=\langle \rho(x)^{-1} \rangle ^{-1}= \frac{d_{\rm A}+d_{\rm B}}{d_{\rm A}\rho_{\rm A}^{-1}+d_{\rm B}\rho_{\rm B}^{-1}}\\
\tilde{\rho}_{\phi}^y&=\tilde{\rho}_{\phi}^z=\langle \rho(x) \rangle = \frac{d_1\rho_{\rm A}+d_{\rm B}\rho_{\rm B}}{d_{\rm A}+d_{\rm B}} \label{rho_y}
\end{align}
Again, the effective parameters arising from the homogenization of the $p$-equation for this one-dimensional example are different from their homogenized $\phi$-equation counterparts. In particular,
\begin{align}
\tilde{\rho}_{p}^x&=\tilde{\rho}_{\phi}^y,\\
\tilde{\rho}_{p}^y&=\tilde{\rho}_{\phi}^x.
\end{align}

\section{Physical nature of the acoustic medium}\label{sec:Analysis}
Both the $p$ and $\phi$ equations are derived from the basic principles of fluid mechanics, which, in the absence of mass sources and external forces are described by the equations~\cite{LANDAU}
\begin{eqnarray}
&&\partial_t \rho + \nabla \cdot (\rho \mathbf{v})=0~, \label{continuity}
\\
&&\rho \mathbf{v}\cdot \nabla \mathbf{v} + \rho \partial_t\mathbf{v} = -\nabla p ~, \label{Euler}
\\
&& \partial_t s + \mathbf{v} \cdot \nabla s=0~, \label{entropy_conservation}
\end{eqnarray}
where $s$ is the entropy. These equations are to be supplemented with the equation of state of the medium, which for the barotropic case can be written as
\begin{equation}\label{state}
p(x^i,t)=g(\rho,x^i),
\end{equation}
where $\rho=f(x^i,t)$ is also a function of space and time. Therefore, in order to find an answer to the above conundrum, we looked at the assumptions required to obtain the $p$ and $\phi$ equations from the previous ones. We will not include here the full derivation of the two equations (the interested reader can consult Refs.~\onlinecite{BAR11-LRR,LANDAU,BER46-JASA,BLO46-JASA}), but only the steps required to understand the physical nature of the medium underlying each of them. In both cases, the first step is the linearization of Eqs.~\eqref{continuity}-\eqref{entropy_conservation}. For that, we express all variables as a sum of a background (subscript 0) and an acoustic (subscript 1) contribution, i.e., $\mathbf{v}=\mathbf{v}_0+\epsilon \mathbf{v}_1$, $p=p_0+\epsilon p_1$, and $\rho=\rho_0+\epsilon \rho_1$, with $\epsilon<<1$. Substitution of these expressions into Eqs.~\eqref{continuity}-\eqref{entropy_conservation} gives rise to a set of equations for the background and acoustic variables, from which the $p$ and $\phi$ equations are derived after imposing some additional conditions. Specifically, to obtain the $p$-equation one must assume that the background or ambient fluid is isobaric~\cite{BER46-JASA}, i.e.,

\begin{equation}
\nabla p_0 = 0.
\end{equation}

On the other hand, to obtain the $\phi$-equation, it is necessary to define an enthalpy function $h$ such that~\cite{BAR11-LRR,LANDAU} (see appendix~\ref{appendix_2})
\begin{align}
\nabla h &= \frac{\nabla p}{\rho},\\
\frac{\partial h}{\partial p} &= \frac{1}{\rho}.
\end{align}
These relations are satisfied when $h$ is only a function of $p$ and the equation of state is of the form $p(x^i,t) = g(\rho)$ or, equivalently,
\begin{equation}\label{globally_barotropic}
\nabla g(\rho_0,x^i)=0.
\end{equation}
Then, the enthalpy can be expressed as
\begin{equation}
h(p) = \int_0^p \frac{1}{\rho(p')}dp'
\end{equation}
We refer to the class of media described by Eq.~\eqref{globally_barotropic} as \emph{globally barotropic}, which are a subset of the more general \emph{locally barotropic} media described by Eq.~\eqref{state}.

\begin{figure}
\begin{center}
  \includegraphics[width=8cm]{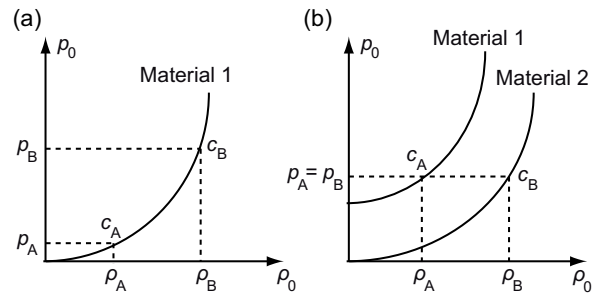}
\end{center}
  \caption{Same set of acoustic parameters arising from two different physical systems. (a) One material at different static pressures (this kind of system is modeled through the $\phi$-equation). (b) Two different materials at the same static pressure (this kind of system is modeled through the $p$-equation).}
  \label{Fig3}
\end{figure}

We can infer more information about the physical systems underlying the $p$ and $\phi$ equations from the equation of state. In particular, it can be shown that, taking the gradient of Eq.~\eqref{state} and linearizing, the corresponding equation for the background reads
\begin{equation}
\nabla p_0 = \frac{\partial g(\rho_0,x)}{\partial \rho_0}\nabla \rho_0+ \nabla g(\rho_0,x^i).
\end{equation}
Thus, the $p$-equation implicitly requires that
\begin{equation}
\nabla \rho_0 = -\frac{\nabla g(\rho_0,x^i)} {\frac{\partial g(\rho_0,x)}{\partial \rho_0}},
\end{equation}
while for the $\phi$-equation we have
\begin{equation}
\nabla \rho_0 = \frac{\nabla p_0}{\frac{\partial g(\rho_0,x)}{\partial \rho_0}},
\end{equation}
As a result,  in the case of the $p$-equation, inhomogeneities in the acoustic parameters are only allowed if they come from having different media at each point (all of them at the same background pressure). This is the usual configuration employed in the construction of metamaterials~\cite{CUM07-NJP,CHE07-APL,TOR08-NJP,NOR08-JASA,CHE10-JPD,POP11-PRL}. However, in the case of the $\phi$-equation, we must have the same medium everywhere, although the acoustic parameters may vary from point to point as a consequence of a background pressure gradient. Therefore, although the values of $\rho_A$ and $\rho_B$, and $c_A$ and $c_B$ in the examples analyzed in section~\ref{sec:Homogenization} are the same in both cases, they represent two entirely different physical systems depending on the considered equation. The situation is outlined in Fig.~\ref{Fig3}. This explains the results of section~\ref{sec:Homogenization}; the considered acoustic parameters represent the wood-air system only when plugged into the $p$-equation, while they represent a fluid subjected to a certain pressure distribution when plugged into the $\phi$-equation.

Since only ATA allows us to work with moving background fluids or implement space-time transformations, the above results are crucial for the understanding of which type of physical systems can actually implement ATA´s prescriptions. Only the inhomogeneities arising from pressure gradients are permitted by the $\phi$-equation around which ATA revolves. Thus, conventional isobaric composites do not qualify for constructing general space-time transformation media with this method. In addition, forcing background pressure gradients in a homogeneous material constitutes an alternative way of building metafluids with properties significantly different from those of isobaric ones. For instance, a remarkable difference that appears in the previous example is the possibility of obtaining supersonic speeds (with respect to the background) for relatively low filling fractions, a feature not displayed by the isobaric composite.

Although the actual attainment of a certain static pressure profile will not be addressed here, we would like to suggest two different potential ways of approaching this issue. The first one is based on the use of one or several pump waves with a much higher amplitude than the acoustic one, which could generate the desired background pressure distribution. The second possibility is related to the transmission of sound in fluids flowing within pipes, where the static pressure depends on the pipe transverse section by Bernoulli's principle. In both scenarios, smooth gradients of the acoustic parameters could be achieved without the need for homogenization. Nonetheless, it is worth pointing out that only the acoustic density can be varied by modifying the background pressure in an ideal gas, since its equation of state is defined by a linear relation. To be able to tailor the speed of sound as well, another more complex material has to be employed, such as the ones governed by polytropic processes. 

Finally, we would like to clarify an additional subtlety that we found with respect to the definition of $c$, where $c=c_0+\epsilon c_1$. This quantity is commonly taken as~\cite{OST05-JASA}
\begin{equation} \label{c_def}
c^2 = ~\frac{\partial p}{\partial \rho}
\end{equation}
A more general definition is given by the relation~\cite{BER46-JASA}
\begin{equation}
 (\partial_t+\mathbf{v}\cdot\nabla)p=c^2(\partial_t+\mathbf{v}\cdot\nabla)\rho
\end{equation}
Both definitions are equivalent for globally barotropic fluids. However, in the locally barotropic case, only the second one remains valid, since it is easy to show that the use of Eq.~\ref{c_def} no longer yields the standard pressure wave equation.

\section{Cloaking the velocity potential}\label{sec:Cloaking}
Using the results of section~\ref{sec:Homogenization} we can now design devices that implement any spatial transformation of the acoustic velocity potential. For this purpose, we express the $\phi$-equation that describes sound propagation in a virtual medium characterized by density $\rho_{\rm V}$ and sound speed $c_{\rm V}$ in arbitrary spatial coordinates $x^i$
\begin{align}
\label{wave_static_arbitrary}
-\frac{\rho_V}{c_{\rm V}^2}\partial^2_t\phi_{1{\rm V}} + \frac{1}{\sqrt{\gamma}}\partial_i\left(\rho_{\rm V}\sqrt{\gamma}\gamma^{ij}\partial_j\phi_{1{\rm V}}\right) = 0.
\end{align}
Under a purely spatial coordinate transformation $\bar{x}^i=f(x^i)$, this equation becomes
\begin{align}
\label{wave_static_arbitrary}
-\frac{\rho_V}{c_{\rm V}^2}\partial^2_t\bar{\phi}_{1{\rm V}} + \frac{1}{\sqrt{\bar{\gamma}}}\partial_i\left(\rho_{\rm V}\sqrt{\bar{\gamma}}\bar{\gamma}^{ij}\partial_j\bar{\phi}_{1{\rm V}}\right) = 0,
\end{align}
where $\bar{\gamma}^{\bar{i}\bar{j}}=\Lambda^{\bar{i}}_i\Lambda^{\bar{j}}_j\gamma^{ij}$, $\Lambda^{\bar{i}}_i$ being the Jacobian of the transformation. In order to mimic the distortion introduced by this transformation we consider a second (real) medium consisting of a microstructure characterized by the position-dependent parameters $\rho_{\rm R}$ and $c_{\rm R}$. As deduced above, sound waves satisfy Eq.~\eqref{eq2} in the long-wavelength regime. This equation can also be expressed in coordinates $x^i$ yielding
\begin{align}
\label{wave_static_homogenized_arbitrary}
-\langle \frac {\rho_{\rm R}} {c_{\rm R}^2} \rangle \partial^2_t\phi_{1{\rm R}} + \frac{1}{\sqrt{\gamma}}\partial_i\left(\sqrt{\gamma}\tilde{\rho}_{\rm R}^{ij}\partial_j\phi_{1{\rm R}}\right)=0,
\end{align}
where the coefficients $\tilde{\rho}_{\rm R}^{ij}$ are obtained via Eq.~\eqref{eq5}. Note that this equation is still valid when $\tilde{\rho}_{\rm R}^{ij}$ has a slow spatial variation (slow spatial variation of the microstructure properties). Clearly, if we rename $\bar{x}^i$ to $x^i$ in Eq.~\eqref{wave_static_arbitrary}, then Eqs.~\eqref{wave_static_arbitrary} and \eqref{wave_static_homogenized_arbitrary} are formally identical if
\begin{align}
\frac{\tilde{\rho}_{\rm R}^{ij}}{\rho_{\rm V}}&=\frac{\sqrt{\bar{\gamma}}}{\sqrt{\gamma}}\bar{\gamma}^{ij} \\
\langle \frac {\rho_{\rm R}} {c_{\rm R}^2} \rangle &= \frac{\sqrt{\bar{\gamma}}}{\sqrt{\gamma}} \frac{\rho_V}{c_{\rm V}^2}
\end{align}
\subsection{Example: cylindrical cloak}
Consider the following radial transformation in cylindrical coordinates~\cite{SCH06-SCI,TOR08-NJP}
\begin{align}
\label{Transformation}
\nonumber \bar{r}&=\frac{b-a}{b}r+a\\ 
		   \bar{\theta}&=\theta\\
\nonumber \bar{z}&=z
\end{align}
According to the previous results, the effective parameters required to implement this transformation are
\begin{align}
\label{Cloak_density}
\rho^{ij}&=\rho_{\rm V}\left(
\begin{array}{ccc}
\frac{r-a}{r} & 0 & 0 \\
0 & \frac{r}{r-a} & 0 \\
0 & 0 & \frac{r-a}{r}\left(\frac{b}{b-a}\right)^2
\end{array}\right)\\
\label{Cloak_coeff}
\langle \frac {\rho_{\rm R}} {c_{\rm R}^2} \rangle &= \frac{r-a}{r}\left(\frac{b}{b-a}\right)^2 \frac{\rho_V}{c_{\rm V}^2}
\end{align}
To implement these parameters, we consider a cylindrical multilayer structure made up of alternating low- and high-density materials characterized by parameters $\rho_{\rm A}$, $c_{\rm A}$ and $\rho_{\rm B}$, $c_{\rm B}$, respectively. Note that a similar strategy was adopted in previous works on electromagnetic and thermal cloaking~\cite{HUA07-OE,QIU09-PRE,GUE12-OE,HAN13-PIER}. Again, the length of each low- (high-) density layer is supposed to be $d_{\rm A}$ ($d_{\rm B}$). To a good approximation, we can use the results of Eqs.~\eqref{rho_x} and ~\eqref{rho_y} to obtain the effective density of this structure by taking $\tilde{\rho}_{\phi}^r=\tilde{\rho}_{\phi}^x$ and $\tilde{\rho}_{\phi}^\theta=\tilde{\rho}_{\phi}^y$. Following the procedure of Ref.~\onlinecite{TOR08-NJP}, we simplify the design by setting the thickness of all layers to a fixed value $d_{\rm A}=d_{\rm B}=d$ and impose the following conditions
\begin{align}
\frac{\rho_{\rm A}}{\rho_{\rm V}}&=\frac{\rho_{\rm V}}{\rho_{\rm B}}\\
c_{\rm A}&=c_{\rm B}
\end{align}

\begin{figure}
\begin{center}
  \includegraphics[width=7cm]{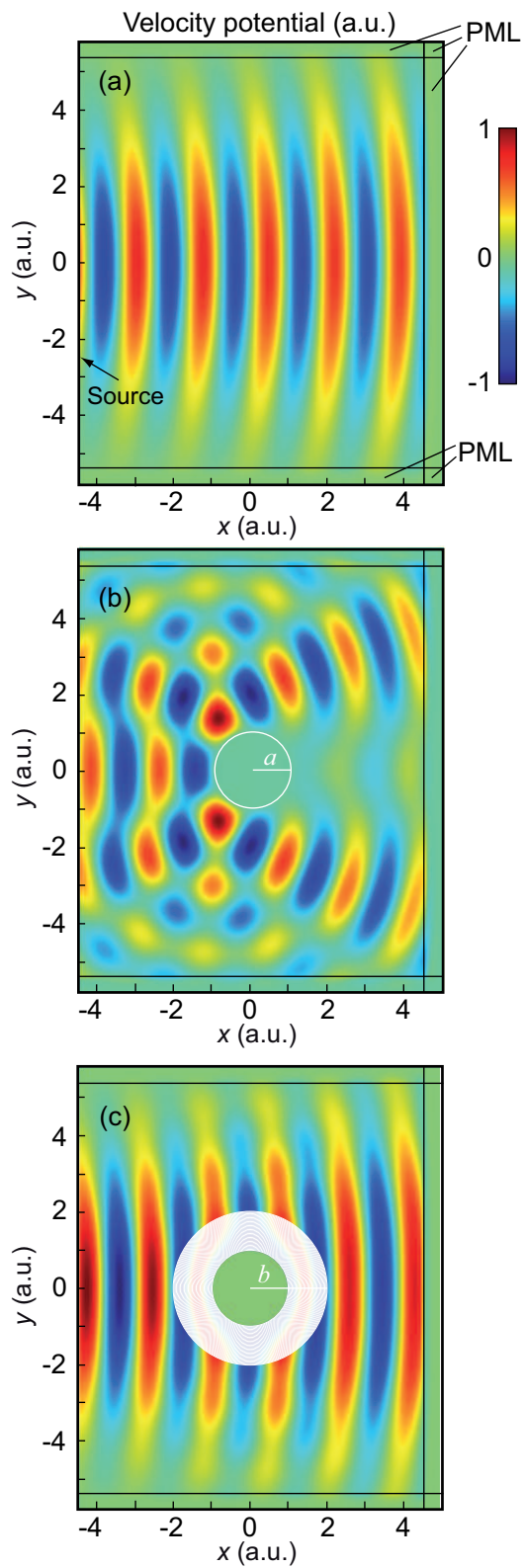}
\end{center}
  \caption{(Color online) Gaussian beam (a) propagating through air, (b) impinging onto a high-density cylinder, (c) and onto the cloak-surrounded cylinder. The cloak consists of 50 different layers with $b=2a=1.7\lambda$, and $d\approx\lambda/100$, where $\lambda$ is the acoustic wavelength. In all simulations, the left boundary is defined as the source of the Gaussian beam. The rest of the computational domain is terminated by PMLs.}
  \label{Fig4}
\end{figure}

As a result, we obtain
\begin{align} 
\tilde{\rho}^r_{\phi}&=\frac{2}{\rho_{\rm A}^{-1}+\rho_{\rm B}^{-1}}\\
\tilde{\rho}^{\theta}_{\phi}&=\langle \rho(r) \rangle = \frac{\rho_{\rm A}+\rho_{\rm B}}{2}\\
\langle \frac {\rho_{\rm R}} {c_{\rm R}^2} \rangle &= \frac {\rho_{\rm A}+\rho_{\rm B}} {2c_{\rm A}^2} 
\end{align}
For these values to be equal to those specified by Eqs.~\eqref{Cloak_density}-\eqref{Cloak_coeff}, it is clear that
\begin{align}
\rho_{\rm A}&=\frac{\rho_{\rm V}}{r-a}\left(r+\sqrt{2ar-a^2} \right)\\
c_{\rm A}&=\frac{r}{r-a}\frac{b-a}{b}c_{\rm V}
\end{align}
To verify the functionality of the designed cloak we solved the velocity potential equation in COMSOL for different situations (see Fig.~\ref{Fig4}).
First, we simulated a Gaussian beam propagating through air. If a high-density cylinder is placed in its way, there appear shadows and reflections. These effects 
are suppressed if the cylinder is surrounded by the designed multilayer. According to our previous analysis, note that the whole cloak should consist of a unique fluid, while the different densities and sound speeds required at each region (layer) should be obtained by forcing the corresponding background pressure.

In addition, note that this cloak implements the transformation given by Eq.~\eqref{Transformation} over the velocity potential. Therefore, the pressure distribution inside the cloak is different from the one existing in a device designed to cloak the pressure, as the one in \onlinecite{TOR08-NJP}. Nevertheless, both devices prevent the acoustic wave from entering the inner region, which is equally cloaked in the two cases.

\section{Conclusions} \label{sec:Conclusions}
In this paper we have applied a homogenization process to the velocity potential acoustic wave equation. This allowed us to derive the actual laboratory realization of acoustic metamaterials exhibiting the effective properties prescribed by ATA. As an example, we designed a multilayer structure able to cloak the acoustic velocity potential. In addition, the analysis of several examples revealed an important difference in the way in which the results of STA and ATA should be implemented in a real device. In particular it appears clear that STA can be used only to design isobaric systems whereas ATA can be used only to design globally barotropic ones. Such conclusion depends critically on the assumptions on the thermodynamical properties of the fluid at the base of the derivation of the acoustic equations and it points towards new experimental approaches to the construction of acoustic metafluids.

\section*{Acknowledgments}
This work was developed under the framework of the ARIADNA contract 4000104572/12/NL/KML of the European Space Agency. C.~G.-M., J.~S.-D., and A.~M. also acknowledge support from Consolider project CSD2008-00066, A.~M. from project TEC2011-28664-C02-02, and C.~B. and G.~J. from the project FIS2011-30145-C03-01. J. S.-D. acknowledges support from the USA Office of Naval Research.

All authors contributed equally to this work.

\appendix
\section{Review of homogenization basics} \label{appendix_1}
To homogenize Eq.~\eqref{eq1} we use a so-called two-scale approach, which studies the asymptotic behavior of the equation as the medium periodicity tends to zero~\cite{Bensoussan,HAS98-CS}. The dependence on the medium periodicity is encoded through a parameter $\epsilon$, which is proportional to its length scale. Then, the coefficients $a^{ij}$ are expressed as a function of $\mathbf{y}=\mathbf{x}/\epsilon$, while any other variable depends both on $\mathbf{x}$ and $\mathbf{y}$. Thus, for each value of $\epsilon$ we have an equation
\begin{align}
\label{inhomogeneous}
-C(\mathbf{y})\partial^2_t\phi_1(\mathbf{x},\mathbf{y},t) + \mathcal{A}^\epsilon \phi_1(\mathbf{x},\mathbf{y},t) = 0.
\end{align}
where we have defined the operator $\mathcal{A}^\epsilon$ as
\begin{align}
\mathcal{A}^\epsilon := \frac{\partial}{\partial x^i}\left(a^{ij}(\mathbf{y})\frac{\partial}{\partial x^j}\right)
\end{align}
Using the chain rule, we can express $\mathcal{A}^\epsilon$ as
\begin{align}
\label{operator}
\mathcal{A}^\epsilon = \epsilon^{-2}\mathcal{A}^0+\epsilon^{-1}\mathcal{A}^1+\mathcal{A}^2
\end{align}
with
\begin{align}
\label{A0} \mathcal{A}^0 &:= \frac{\partial}{\partial y^i}\left(a^{ij}(\mathbf{y})\frac{\partial}{\partial y^j}\right)\\
\label{A1} \mathcal{A}^1 &:= \frac{\partial}{\partial x^i}\left(a^{ij}(\mathbf{y})\frac{\partial}{\partial y^j}\right)+\frac{\partial}{\partial y^i}\left(a^{ij}(\mathbf{y})\frac{\partial}{\partial x^j}\right)\\
\label{A2} \mathcal{A}^2 &:= \frac{\partial}{\partial x^i}\left(a^{ij}(\mathbf{y})\frac{\partial}{\partial x^j}\right)
\end{align}
We seek a solution of the form
\begin{align}
\label{solution}
\phi_1(\mathbf{x},\mathbf{y},t)=\phi_1^0(\mathbf{x},\mathbf{y},t)+\epsilon\phi_1^1(\mathbf{x},\mathbf{y},t)+\epsilon^2\phi_1^2(\mathbf{x},\mathbf{y},t),
\end{align}
where each function $\phi_1^i$ is periodic in $\mathbf{y}$. Substituting Eq.~\eqref{operator} and Eq.~\eqref{solution} into Eq.~\eqref{inhomogeneous} and equating equal powers of $\epsilon$ we obtain the following set of equations
\begin{align}
\mathcal{A}^0\phi_1^0&=0 \label{Equation_1} \\
\mathcal{A}^0\phi_1^1&=-\mathcal{A}^1\phi_1^0 \label{Equation_2} \\
\mathcal{A}^0\phi_1^2&=C(\mathbf{y})\partial^2_t\phi_1^0 - \mathcal{A}^1\phi_1^1 - \mathcal{A}^2\phi_1^0 \label{Equation_3}
\end{align}
Next, we use a theorem stating that for any $\mathbf{Y}$-periodic function $\phi$, the equation
\begin{align}
\mathcal{A}^0\phi=F
\end{align}
has a (unique) solution (up to a constant) if and only if $\langle F \rangle = 0$~\cite{Bensoussan,HAS98-CS}. As a consequence, we know that there exists a solution to Eq.~\eqref{Equation_1}. 
Moreover, from Eq.~\eqref{Equation_1}, it follows immediately that 
\begin{align} 
\int_Y \phi_1^0\mathcal{A}^0\phi_1^0 d\mathbf{y} = 0
\end{align}
On the other hand, integrating by parts over a unit cell we have
\begin{align} 
\int_Y \phi_1^0\mathcal{A}^0\phi_1^0 d\mathbf{y} + \int_Y a^{ij}\frac{\partial \phi_1^0}{\partial y_i}\frac{\partial \phi_1^0}{\partial y^j} d\mathbf{y} \\ \nonumber
= \int_Y \frac{\partial}{\partial y_i}\left(\phi_1^0 a^{ij} \frac{\partial \phi_1^0}{\partial y^j}\right) d\mathbf{y} = 0,
\end{align}
where the last equality follows from the fact that, if a function $F(\mathbf{y})$ is periodic in $y^i$, then $\int_Y \frac{\partial F}{\partial y^i}d\mathbf{y}=0$ due to the fundamental theorem of calculus. Therefore, we have
\begin{align} 
\int_Y a^{ij}\frac{\partial \phi_1^0}{\partial y_i}\frac{\partial \phi_1^0}{\partial y^j} d\mathbf{y} = 0
\end{align}
Finally, the ellipticity condition allows us to write
\begin{align} 
\int_Y a^{ij}\frac{\partial \phi_1^0}{\partial y_i}\frac{\partial \phi_1^0}{\partial y^j} d\mathbf{y} \ge \alpha\int_Y|\nabla_y \phi_1^0|^2 d\mathbf{y}
\end{align}
Fulfillment of the last two equations implies
\begin{align} 
\nabla_y \phi_1^0 = 0 \rightarrow \phi_1^0(\mathbf{x},\mathbf{y}) = \phi_1^0(\mathbf{x}).
\end{align}
Using this result we obtain
\begin{align}
\mathcal{A}^1\phi_1^0 &= \frac{\partial}{\partial x^i}\left(a^{ij}(\mathbf{y})\frac{\partial \phi_1^0(\mathbf{x})}{\partial y^j}\right)
+\frac{\partial}{\partial y^i}\left(a^{ij}(\mathbf{y})\frac{\partial \phi_1^0(\mathbf{x})}{\partial x^j}\right) \\ \nonumber
&= \frac{\partial\phi_0(\mathbf{x})}{\partial x^j}\frac{\partial a^{ij}(\mathbf{y})}{\partial y^i}.
\end{align}
Thus, $\langle \mathcal{A}_1\phi_1^0 \rangle = 0$ and Eq.~\eqref{Equation_2} has a unique solution, which we assume to be of the form
\begin{align}\label{Solution_2}
\phi_1^1(\mathbf{x},\mathbf{y},t)=\chi^j(\mathbf{y})\frac{\partial \phi_1^0(\mathbf{x},t)}{\partial x^j}+\phi_{\rm C}(\mathbf{x}).
\end{align}
Substitution of Eq.~\eqref{Solution_2} into Eq.~\eqref{Equation_2} gives rise to the so-called \emph{cell problem}
\begin{align}
\label{Cell_problem}
\mathcal{A}^0\chi^k(\mathbf{y})=-\frac{\partial a^{ik}}{\partial y^i},
\end{align}
which provides the sought functions $\chi^j(\mathbf{y})$. The cell problem is guaranteed to have a unique solution, since
\begin{align}
\langle \frac{\partial a^{ik}}{\partial y^i} \rangle=0.
\end{align}
Finally, the solvability condition for Eq.~\eqref{Equation_3} reads
\begin{align}
\label{Solvability_3}
\int_Y \left(\mathcal{A}^1\phi_1^1+\mathcal{A}^2\phi_1^0\right)d\mathbf{y} = \langle C(\mathbf{y}) \rangle \partial^2_t\phi_1^0,
\end{align}
With the help of the following partial results
\begin{align}
\int_Y \mathcal{A}^2\phi_1^0 d\mathbf{y} &= \frac{\partial^2 \phi_1^0(\mathbf{x})}{\partial x^i \partial x^j}\int_Y a^{ij}(\mathbf{y}) d\mathbf{y}\\
\int_Y \mathcal{A}^1\phi_1^1 d\mathbf{y} &= \frac{\partial^2 \phi_1^0(\mathbf{x})}{\partial x^i \partial x^k}\int_Y a^{ij}(\mathbf{y}) \frac{\partial \chi^k(\mathbf{y})}{\partial y^j} d\mathbf{y}
\end{align}
we obtain the \emph{homogenized equation}
\begin{align}
\label{wave_static_homogenized}
-\langle C(\mathbf{y}) \rangle \partial^2_t\phi_1^0 + \tilde{a}^{ij}\frac{\partial^2\phi_1(\mathbf{x})}{\partial x^i\partial x^j}=0,
\end{align}
where the \emph{effective coefficient} $\tilde{a}^{ij}$ is given by
\begin{align}
\label{Effective_density}
\tilde{a}^{ij}=\left\langle a^{ij}+a^{ik}\frac{\partial \chi^j(\mathbf{y})}{\partial y^k}\right\rangle
\end{align}
Note that we have renamed $\mathbf{y}$ to $\mathbf{x}$ in the corresponding equations of the main text to keep the notation simple. 

There exist specific situations for which the cell problem has an analytical solution. One of them is that discussed above, in which the medium properties vary only along the $y_1$ direction, i.e.,
\begin{align}
a^{ij}(\mathbf{y})=a_0(y_1)\delta^{ij}
\end{align}
Let us first analyze the case for $k=1$. Assuming that $\chi^1=\chi^1(y_1)$, the corresponding cell problem is
\begin{align} \label{k_1}
\frac{\partial}{\partial y_1}\left(a_0(y_1)\frac{\partial \chi^1(y_1)}{\partial y_1}\right) = - \frac{\partial a_0(y_1)}{\partial y_1}
\end{align}
Integrating over $y_1$ we obtain
\begin{align}
\frac{\partial \chi^1(y_1)}{\partial y_1} = -1+\frac{K_1}{a_0(y_1)}.
\end{align}
Integrating again
\begin{align}
\chi^1(y_1)=-y_1+K_1\int_0^{y_1}\frac{1}{a_0(y_1)}dy_1 + K_2,
\end{align}
where $K_1$ and $K_2$ are constants. The value of $K_1$ can be determined by using the fact that $\chi^1$ is $Y_1$-periodic (i.e., $\chi(Y_1)-\chi(0)=0$), obtaining
\begin{align}
K_1=\left(\frac{1}{Y_1}\int_0^{Y_1}\frac{1}{a_0(y_1)}dy_1\right)^{-1}= \langle a_0(y_1)^{-1}\rangle^{-1}
\end{align}
It is not necessary to calculate $K_2$, as only the derivatives of $\chi^1$ enter the expression of $\tilde{a}^{ij}$. Moreover, by the theorem above, any other solution will differ from this one by a constant.

For $k = 2$, we have
\begin{align} \label{k_2}
\mathcal{A}^0 \chi^2(\mathbf(y)) = 0 \rightarrow \chi^2={\rm constant}
\end{align}
The same result is obtained for $k=3$. Introducing the calculated functions $\chi^k$ into Eq.~\eqref{Effective_density} we find that
\begin{align} 
\tilde{a}^{ij}&=\langle a_0(y_1) \rangle \delta^{ij} + \left( -\langle a_0(y_1) \rangle + \langle a_0(y_1)^{-1} \rangle ^{-1} \right) \delta^{1i}\delta^{1j}.
\end{align}
Therefore,
\begin{align}
\tilde{a}^{ij}=\left(
\begin{array}{ccc}
\langle a_0(y_1)^{-1} \rangle^{-1} & 0 & 0 \\
0 & \langle a_0(y_1) \rangle & 0 \\
0 & 0 & \langle a_0(y_1) \rangle.
\end{array}
\right)
\end{align}
%
\section{Derivation of the velocity potential equation} \label{appendix_2}
We start from the basic equations of fluid mechanics, Eqs.~\eqref{continuity}-\eqref{entropy_conservation}. First, we use the identity
\begin{align}
\label{Identity}
\left(\mathbf{v}\cdot\nabla\right)\mathbf{v}=\nabla\left(\frac{1}{2}\mathbf{v}^2\right)-\mathbf{v}\times\left(\nabla\times\mathbf{v}\right),
\end{align}
to transform Euler's equation [Eq. \eqref{Euler}] to
\begin{align}
\label{Euler2}
\partial_t \mathbf{v}=\mathbf{v}\times\left(\nabla\times\mathbf{v}\right)-\frac{1}{\rho}\nabla{p}-\nabla\left(\frac{1}{2}v^2\right).
\end{align}
Using the definition for the velocity potential ($\mathbf{v}=-\nabla \phi$), the last equation reduces to
\begin{align}
\label{Euler3}
\partial_t \mathbf{v}=-\frac{1}{\rho}\nabla{p}-\nabla\left(\frac{1}{2}v^2\right).
\end{align}
If the fluid moreover is globally barotropic [$\rho=\rho\left(p\right)$], we can define the enthalpy as
\begin{align}
\label{enthalpy}
h\left(p\right)=\int_{0}^{p}\frac{dp'}{\rho\left(p'\right)},
\end{align}
and considering that
\begin{align}
\frac{dh\left(p\right)}{dp}=\frac{d}{dp}\int_{0}^{p}\frac{dp'}{\rho\left(p'\right)}=\frac{1}{\rho\left(p\right)},
\end{align}
Eq.~\eqref{enthalpy} implies that
\begin{align}
\nabla{h\left(p\right)}=\frac{dh\left(p\right)}{dp}\nabla{p}=\frac{1}{\rho\left(p\right)}\nabla{p}.
\end{align}
As a consequence, Eq.~\eqref{Euler3} can be expressed as
\begin{align}
\label{Euler4}
\partial_t \mathbf{v}=-\nabla{h}-\nabla\left(\frac{1}{2}v^2\right).
\end{align}
Using again the relation $\mathbf{v}=-\nabla\phi$, Euler's equation finally becomes
\begin{align}
\label{Euler5}
-\partial_t \phi+h+\frac{1}{2}\left(\nabla\phi\right)^2=0.
\end{align}
Now we proceed to linearize the equation of continuity, Eq.~\eqref{continuity}, and this last version of Euler's equation.
As usual, we express the variables involved in these equations as $\rho=\rho_0+\epsilon\rho_1$, $p=p_0+\epsilon{p}_1$, and
$\phi=\phi_0+\epsilon\phi_1$, where the subscripts 0 and 1 indicate ambient values (in the absence of acoustic perturbations)
and their fluctuations (due to a propagating acoustic wave), respectively. We introduce a similar definition for the velocity based
on the linearization of the potential
\begin{align}
\label{Lin_velocity}
-\nabla{\phi}=-\nabla{\phi_0}+\epsilon\left(-\nabla\phi_1\right) \Rightarrow \mathbf{v}=\mathbf{v_0}+\epsilon\mathbf{v_1}.
\end{align}
Inserting these linearized variables into the continuity equation yields
\begin{align}
\nonumber
\partial_t \rho_0&+\epsilon\partial_t \rho_1+\nabla\cdot\left(\rho_0\mathbf{v_0}\right)+\epsilon\nabla\cdot\left(\rho_0\mathbf{v_1}+\rho_1\mathbf{v_0}\right) \\ 
&+\epsilon^2\nabla\cdot\left(\rho_1\mathbf{v_1}\right)=0. \label{Lin_continuity}
\end{align}
Neglecting second-order terms, we obtain the following equation for the (first-order) acoustic perturbation
\begin{align}
\partial_t \rho_1+\nabla\cdot\left(\rho_0\mathbf{v_1}+\rho_1\mathbf{v_0}\right)=0. \label{Lin_continuity2}
\end{align}
Now we use the first order Taylor expansion of $h\left(p\right)$
\begin{align}
\nonumber
h\left(p\right)&=h\left(p_0+\epsilon{p}_1\right)=h\left(p_0\right)+\left(\epsilon{p}_1\right)\frac{\partial{h}\left(p\right)}{\partial{p}}\bigg|_{p=p_0} \\
&=h_0+\left(\epsilon{p}_1\right)\frac{1}{\rho\left(p\right)}\bigg|_{p=p_0}=h_0+\epsilon\frac{{p}_1}{\rho_0}, \label{Taylor_h}
\end{align}
to linearize Eq.~\eqref{Euler5}, which becomes
\begin{align}
\nonumber
-\partial_t \phi_0-\epsilon\partial_t \phi_1+h_0+\epsilon\frac{{p}_1}{\rho_0}+\frac{1}{2}\left(\nabla\phi_0\right)^2 \\
+\epsilon \nabla\phi_0 \cdot\nabla\phi_1+\epsilon^2\frac{1}{2}\left(\nabla\phi_1\right)^2=0. \label{Lin_Euler}
\end{align}
Again, neglecting second-order terms, we arrive at the following acoustic equation
\begin{align}
p_1=\rho_0\left(\partial_t \phi_1+\mathbf{v_0}\cdot\nabla\phi_1\right).  \label{Lin_Euler2}
\end{align}
From the expansions of $\rho$ and $p$, we obtain the relation
\begin{align}
\label{Bulk}
\rho_1=\frac{d\rho_0}{dp_0}p_1~=\frac{1}{c_0^2}p_1.
\end{align}
Insertion of Eq.~\eqref{Lin_Euler2} into Eq.~\eqref{Bulk} leads to
\begin{align}
\label{rho1}
\rho_1=c_0^{-2}\rho_0\left(\partial_t \phi_1+\mathbf{v_0}\cdot\nabla\phi_1\right).
\end{align}
Finally, we substitute Eq.~\eqref{rho1} into the linearized equation of continuity, Eq.~\eqref{Lin_continuity2}, to
obtain a wave equation for the velocity potential
\begin{eqnarray}
&-&\partial_t\left[\rho_0{c_0}^{-2}\left(\partial_t\phi_1 +\mathbf{v}_0\cdot\nabla\phi_1\right)\right] \\ \nonumber
&+&\nabla\cdot\left[\rho_0\nabla\phi_1-\rho_0 c_0^{-2}\left(\partial_t\phi_1 + \mathbf{v}_0\cdot\nabla\phi_1 \right)\mathbf{v}_0\right]=0.
\end{eqnarray}
In the case in which $\mathbf{v}_0=0$ and the acoustic parameters do not depend on time (the case studied in this work), the above equation reduces to
\begin{align} 
\label{wave_no_moving}
-\rho_0 c_0^{-2} \partial_t^2 \phi_1+\nabla\cdot \left( \rho_0\nabla\phi_1 \right)=0.
\end{align}

\end{document}